\documentstyle[12pt,twoside]{article}

\normalbaselines
\font\tbf = cmbx12

\pagebreak

\begin{document}

\indent
\vskip 0.5cm
\centerline{\tbf NON-COMMUTATIVE  EXTENSIONS}
\vskip 0.3cm
\centerline{\tbf OF  CLASSICAL  THEORIES  IN  PHYSICS}
\vskip 1.4cm
\centerline{\tbf Richard Kerner} 
\vskip 0.5cm
\centerline{\it Laboratoire de Gravitation et Cosmologie Relativistes,}
\vskip 0.3cm
\centerline{\it Universit\'e Pierre-et-Marie-Curie - CNRS ESA 7065,} 
\vskip 0.3cm
\centerline{\it Tour 22, 4-\`eme \'etage, Bo\^{i}te 142} 
\vskip 0.3cm
\centerline{\it 4, Place Jussieu, 75005 Paris, France}
\vskip 0.2cm
\centerline{{\it e-mail}: \ \ rk@ccr.jussieu.fr}
\vskip 0.8cm
\indent
\hskip 0.2cm
{\tbf Abstract} \, {\small We propose a short introductory overview of the
non-commutative extensions of several classical physical theories. After 
a general discussion of the reasons that suggest that the non-commutativity
is a major issue that will eventually lead to the unification of gravity
with other fundamental interactions, we display examples of non-commutative
generalizations of known geometries. 
\newline
\indent
Finally we discuss the general properties of the algebras that could become
generalizations of algebras of smooth functions on Minkowskian (Riemannian)
manifolds, needed for the description of Quantum Gravity.}
\newpage
\indent
{\tbf 1. Deformations of Space-Time and Phase space Geometries.}
\vskip 0.3cm
\indent
The two most important branches of modern physics created in the beginning
of this century, the General Relativity and Quantum Theory, possess their
well-defined classical counterparts, the Newtonian gravity theory mechanics, 
which are obtained as limits of these theories when the parameters $c^{-1}$ 
or $\hbar$ The mathematical expression of this fact is formulated in terms of 
the {\it deformations} of the respective structures. The notion of deformation 
plays the central r\^ole in modern attempts which try to generalize the 
geometrical description of physical realm.
\newline
\indent
To be more precise, we can cite the example of the relation existing between 
the Lorentz and the Galilei groups: the Lorentz group can be considered as 
deformation of the Galilei group, with the characteristic parameter $c^{-1}$; 
when this parameter tends to zero, the Lorentz group is said to undergo the 
{\it contraction} into the Galilei group. Similarly, the quantization 
procedure proposed by J.E.Moyal [1] is a deformation of the usual Poisson 
algebra which is contracted back to it when the characteristic parameter of 
deformation which is here the Planck constant $h $ tends to zero. Finally, 
Special Relativity may be considered as a contraction of General Relativity 
when the characteristic parameter $G $ tends to zero (although some 
space-times different from the Minkowskian one can appear when the Ricci 
tensor is put to zero).
\newline
\indent
Now, with three fundamental constants of Nature, $h$, $G$ and $c^{-1}$
serving as deformation parameters, one can imagine seven different
contractions of the hypothetical unified theory that would deserve
the name of ``Relativistic Quantum Gravity'', and which is yet to
be invented.
The seven contractions correspond to the vanishing of:
\vskip 0.2cm
\vskip 0.3cm
a) one of the three parameters, i.e $h$, $G$, or $c^{-1}$ only;
\newline 
\indent
b) two parameters at once, i.e.($h$ and $G$), ($h$ and $c^{-1}$), and 
($G$ and $c^{-1}$);
\newline
\indent
c) all the three parameters at once, i.e. (${h}$, ${G}$ and $c^{-1}$).
\vskip 0.5cm
The following Table shows the relations between the corresponding theories, 
as well as their usual denominations (when we know them...). We did not take 
into account the fact that taking the double limits might be non-commutative, 
which cannot be excluded {\it a priori} and would have made our diagram even
more complicated.
\newline
\indent
Two of the theories displayed here have not found their realization yet: 
the ``Relativistic Quantum Gravity'' and the ``Non-Relativistic Quantum 
Gravity''. It is not at all clear whether these hypothetical theories can be 
realized without introducing some new deformation parameter depending on a 
new physical constant, and whether this constant should be independent or 
related to the three fundamental constants $h $, $c $ and $G $ or not.
\newline
\indent
It is also amusing to note that our diagram is three-dimensional - is it
just a coincidence that we happen to live in three space dimensions, too ?

\newpage
\indent
\vskip 0.4cm

%Diagram
% Here you specify your preferred unit length and the thickness
% of your lines

\unitlength=1mm       

\linethickness{1pt}                      

% all dimensionless numbers specifying lengths are mm
% this is a relatively thick line

\def\sh{\strut \\}

% Here, introduce some geometric constants, such as the size of the
%  "pudlo" box, the vertical distance between pudlos etc. 

\def\pudwidth{40}
\def\pudheight{20}
\def\between{20}

\def\pudlo#1{\framebox(\pudwidth,\pudheight)[cc]{\shortstack{#1}}}
\def\vlen{\between}
\def\vs{\vector(-1,-1){\vlen}}
\def\vl{\vector(-1,0){\vlen}}
\def\vd{\vector(0,-1){\vlen}}

% Levels of the first three rows. The fourth is placed at y=0. Adapted
%  to the vertical size=140 mm.
\def\frst{122}
\def\scnd{78}
\def\thrd{47}

% Our favourite limits
\def\limh{\makebox(10,6){$\hbar \to 0$}}
\def\limg{\makebox(10,6){$G \to 0$}}
\def\limc{\makebox(10,6){$c^{-1} \to 0$}}

\noindent

\begin{picture}(160,140)

% First row

\put(24,\frst){\pudlo{\it ``Non-Relativistic\sh \it Quantum\sh \it Gravity''}}
\put(92,\frst){\pudlo{\it ``Relativistic\sh \it Quantum\sh \it Gravity''}}

% Now the arrows

\put(13,47){\vd}
\put(13,75){\vd}
\put(78,47){\vd}
\put(78,75){\vd}
\put(53,88){\vd}
\put(53,119){\vd}
\put(120,88){\vd}
\put(120,119){\vd}

\put(63,15){\vl}
\put(63,90){\vl}
\put(88,57){\vl}
\put(88,130){\vl}

\put(45,45){\vs}
\put(45,120){\vs}
\put(112,45){\vs}
\put(112,120){\vs}

% ... and their description

\put(33,35){\limh} 
\put(100,35){\limh}
\put(33,110){\limh}
\put(100,110){\limh}

\put(08,48){\limg} 
\put(73,48){\limg}
\put(115,90){\limg}
\put(48,105){\limg}

\put(49,15){\limc}
\put(49,90){\limc}
\put(74,57){\limc}
\put(74,130){\limc}

% Next row

\put(0,\scnd){\pudlo{Newton's Gravity\sh and\sh Mechanics}}
\put(68,\scnd){\pudlo{General\sh Relativity}}

% Next row

\put(24,\thrd){\pudlo{Non-relativistic\sh Quantum\sh Mechanics}}
\put(92,\thrd){\pudlo{Relativistic\sh Quantum Field\sh Theory}}

% Bottom row

\put(0,4){\pudlo{Classical Mechanics\sh without gravity}}
\put(68,4){\pudlo{Special\sh Relativity}}

\end{picture}

\vskip 0.2cm
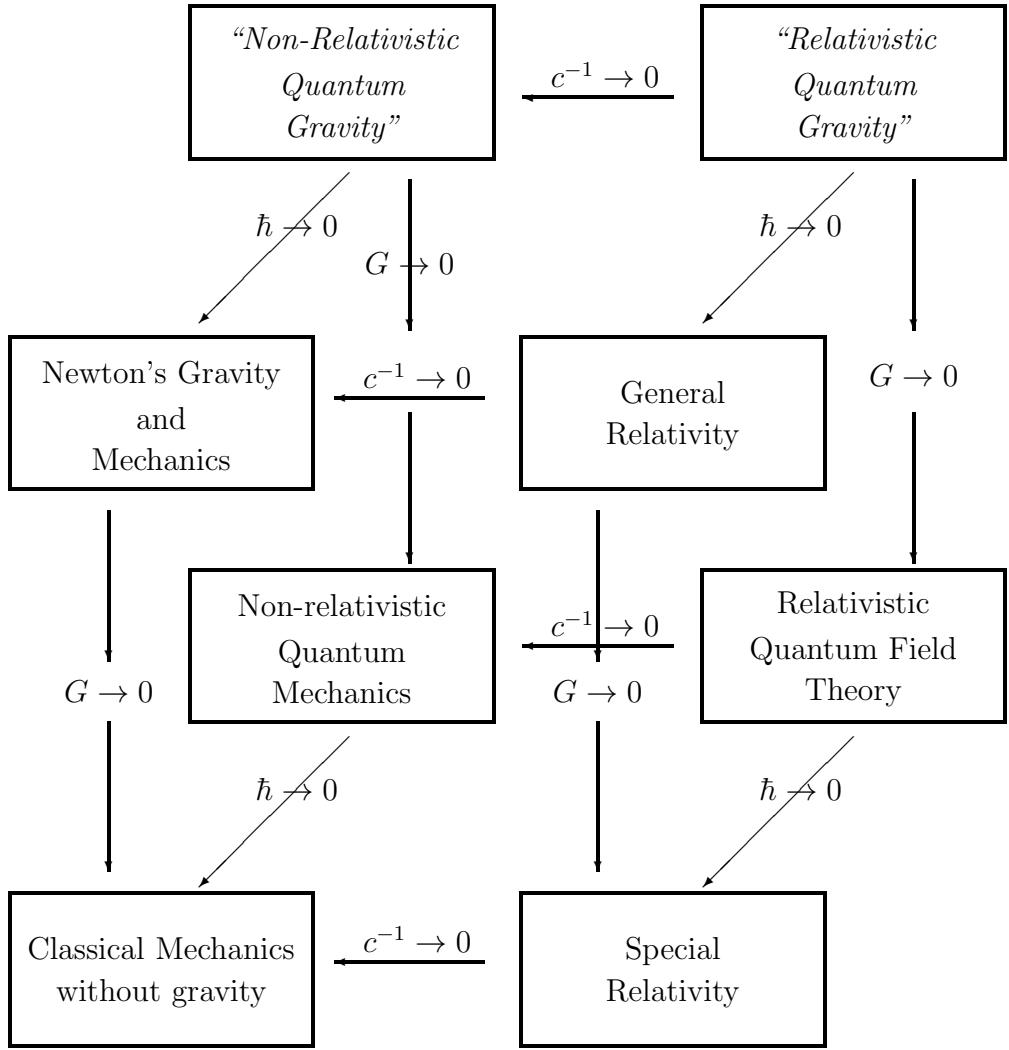
\centerline{ Eight limits of fundamental physical theories}
\vskip 0.1cm
\centerline{\it Two limits (marked in italics) are still to be invented}
\vskip 0.3cm
\indent
In the above figure, the contractions (symbolized by the arrows coinciding
with the edges of the cube) relate two-by-two different space-time or phase
space geometries. The best way to describe a {\it geometry} is, in our sense,
to define the set of variables (forming an algebra) that in a natural way 
would generalize the algebra of local coordinates in these spaces.
\newline
\indent
P.A.M.Dirac was already aware of the possibility of a radical modification 
of geometrical notions, and in his fundamental papers written in 1926 $[2]$ 
he evokes the possibility of describing the phase space physics in terms of 
a non-commutative analogue of the algebra of functions, which he referred 
to as the ``quantum algebra'', together with its derivations, which he 
called ``quantum differentiations''. 
\newpage
\indent
Of course, this kind of geometry seemed strange and even useless
from the point of view of General Relativity. Einstein thought that
further problems of physics should be solved by subsequent development
of geometrical ideas, and it seemed to him that to have $a\times b $ not
equal to $b\times a $ was something that does not fit very well with 
geometry as he understood it [3]
\newline
\indent
During several decades, mostly in the sixties and the seventies,
a lot of efforts have been made in order to find a unifying approach
to both these great theories.In doing so, people either tried to 
generalize one of the two theories so that the other one would follow, or 
tried to merge them together via embedding into some more general unified 
theory. Most of the activities in this field rather belonged to the first 
category. 
\newline
\indent
The Hamiltonian formulation of General Relativity by R.Arnowitt, S.Deser and
C.Misner [4], and later the Wheeler--De Witt equation which generalizes
Schr\"odinger's equation for quantum wave functions describing the state of
a 3-dimensional geometry of the Universe [5] can be considered as a first
attempt to quantize the General Relativity. The geometric quantization
developed by J.M.Souriau, D.Simms, and B.Kostant ([6], [7], [8]) tried to
derive the rules of quantum mechanics by interpreting the observables and
state vectors as elements of algebras of operators and functions defined on
classical manifolds with sufficiently rich geometry, (e.g. symplectic
manifolds, fibre bundles, jet spaces).
\newline
\indent
Simultaneous consideration of the two most important new physical theories
of this century, the General Relativity and Quantum Mechanics, did not
bring a common tool for the description of the nature of spacetime at the
microscopic level. The General Relativity develops our knowledge about global
properties of space and time at very large distances, and raises
the questions concerning the global topology of the Universe. 
\newline
\indent
The methods of Differential Geometry which are the best adapted 
as the mathematical language of this theory, are very different from the 
methods of Quantum Physics, in which one studies the properties of the 
algebra of observables, considering the state vectors, as well as geometric 
points and trajectories, as artefacts and secondary notions. This approach 
has been inspired by the works of John von Neumann [9], and has much in 
common with the non-commutative geometry, where the very notion of a point 
loses its meaning.
\newline
\indent
A strong flavor of non-commutativity is also present in A. Ashtekar's approach
to quantum gravity, in which the notion of coordinates becomes secondary, the
only intrinsic information being encoded in the loop space (see, e.g. in
A. Ashtekar [10], or C. Rovelli [11])
\newline
\indent
In the next section, we shall give a few arguments in favor of the hypothesis
that the realization of a theory taking into account quantum effects in
gravitation should also lead to the abandon of usual notion of coordinates
and differential manifolds and to the introduction of non-commutative 
extensions of algebras of smooth functions on manifolds. We shall also see 
that such algebras can act on free modules, which becomes a natural 
generalization of gauge theories described mathematically as connections 
and curvatures on fibre bundles.
\newpage
\indent
{\tbf 2. Why the coordinates should not commute at Planck's scale.}
\vskip 0.3cm
\indent
There are several well-known arguments which suggest that the dynamical 
interplay between Quantum Theory and Gravitation should lead to a 
non-commutative version of space-time. Let us recall the few ones that are
cited most frequently:
\vskip 0.2cm
\indent
\hskip 0.5cm
* A semi-classical argument that involves black-hole creation at very small
distances: as a matter of fact, if the General Relativity remains valid at
the Planck scale, then any localization of events should become impossible
at the distances of the order of ${\lambda}_P = \sqrt{\frac{\hbar G}{c^3}}$.
Indeed, according to quantum mechanical principles, lo localize an event
in space-time within the radius $\Delta \, x^{\mu} \sim a$, one need to
employ the energy of the order $a^{-1}$. When $a$ becomes too small, the
creation of a mini black hole becomes possible, thus excluding from the
observation that portion of the space-time and making further localization
meaningless.
\newline
\indent
Therefore, the localization is possible only if we impose the following 
limitation on the time interval:
\begin{equation}
\Delta \, x^0 \, ( \Sigma \, \Delta \, x^k) \, \geq {\lambda}^2_P \, \ \ \, 
{\rm and} \, \ \ \, \ \ \Delta \, x^k \, \Delta \, x^m \geq {\lambda}^2_P .
\end{equation}
in order to avoid the black hole creation at the microscopic level.
\vskip 0.2cm
\indent
\hskip 0.5cm
** The topology of the space-time should be sensitive to the states of the 
fields which are in presence - and {\it vice versa}, quantum evolution of
any field, including gravity, should take into account all possible field
configurations, also corresponding to the fields existing in space-times
with radically different topologies (a creation of a black hole is but the
simplest example; one should also take into account other ``exotic''
configurations, such as multiple Einstein-Rosen bridges (the so-called
``{\it wormholes}'', leading in the limit of great $N$ to the space-time 
{\it foam}. 
\newline
\indent
Now, as any quantum measurement process may also lead to topological 
modifications, again the coordinates of an event found before and after any
measurement can no more be compared, because they might refer to uncompatible
coordinate patches in different local maps. As a result, quantum measures
of coordinates themselves become non-commutative, and the algebra of functions
on the space-time, supposed to contain also all possible local coordinates,
must be replaced by its non-commutative extension, better adapted to describe
the space-time foam.
\vskip 0.2cm
\indent
\hskip 0.5cm
*** Since the coordinates $x^{\mu}$ are endowed with a length scale, the
metric must enter at certain stage in order to measure it. After quantization,
the components of the tensor $g_{\mu \nu}$ become a set of dynamical fields, 
whose behaviour is determined by the propagators and, at least at the lowest
perturbative level, by two-point correlation functions. As any other field,
the components of the metric tensor will display quantum fluctuations, making
impossible precise measurements of distances, and therefore, any precise
definition of coordinates. 
\newpage
\indent
Our aim here is not to discuss all possible arguments suggesting that at the 
Planck scale not only the positions and momenta do not commute anymore, but
also the coordinates themselves should belong to a non-commutative algebra.
In what follows, we shall take it for granted that such is the case, and
shall expose in a concise way, on the example of the simplest finite
non-commutative algebra, which is the algebra of complex $n \times n$
matrices, how almost all the notions of usual differential geometry can be
extended to the non-commutative case. We shall also show how the gauge
theories and the analogs of the fibre bundle spaces and Kaluza-Klein 
geometries can be generalized in the non-commutative setting.
\newline
\indent
Finally, as our main subject is the hypothetical Quantum Gravity theory, and
because it has to have also a limit as Relativistic Field theory when gravity
is switched off, we shall analyze the conequences of the Poincar\'e invariance
that must be imposed on any theory of this type.
\vskip 0.4cm
\indent
{\tbf 3. Non-commutative differential geometry}
\vskip 0.3cm
\indent
In the examples of non-commutative generalizations of spaces of states or of 
algebras of observables, we have looked up to now only at the linear cases. 
A most general non-commutative geometry should imitate the situations 
encountered in the ordinary differential geometry of manifolds. Therefore, 
we should replace the algebra of smooth functions on a manifold, (the maximal 
ideals of this algebra can be identified as points of the corresponding 
manifold) by an more general associative algebra, which can be 
non-commutative. The {\it derivations} of this algebra will naturally 
generalize the notion of {\it vector fields}; their dual space will 
generalize the fields of {\it 1-forms}, and one can continue as far as 
possible, trying to construct the analogues of a metric, integration, volume 
element, Hodge duality, Lie derivatives, connection and curvature, and so 
forth. It is amazing how almost all of these objects known from the classical
version of differential geometry find their counterparts in the 
non-commutative case.
\newline
\indent
The differential algebras of this type have been studied by A.Quillen [12], 
A.Connes [13] and M.Dubois-Violette [14]; their application to new 
mathematical models of the gauge theories, including the standard model 
of electroweak interactions of Weinberg and Salam, has been worked out by 
M.Dubois-Violette {\it et al} [15],[16], by A.Connes and J.Lott [17], 
R.Coquereaux {\it et al} [18], and many other authors since. Here we shall 
give the simplest example of realization of the non-commutative geometry 
proposed in [15],[16], realized with the algebra of complex $n\times n $ 
matrices, $M_n $({\tbf C}). Any element of $M_n $({\tbf C}) can be 
represented as a linear combination of the unit $n\times n $ matrix 
{\tbf 1} and $(n^2 - 1) $ hermitian traceless matrices 
$E_k $, k = 1,2, \dots  ,$(n^2 - 1) $:
\begin{equation}
B = \beta \ \ {\tbf 1 } + \sum{\alpha^k E_k}
\end{equation}
One can choose the basis in which the following relations hold:
\begin{equation}
{E_k}{E_m} = ({1\over n})g_{km} {\tbf 1} + S_{km}^{j} E_j - ({i\over 2})
C_{km}^{j} E_j
\end{equation}
with real coefficients satisfying $S_{km}^j $ = $S_{mk}^j $,
$S_{km}^k = 0 $, $C_{km}^j = - C_{mk}^j $, $C_{km}^k = 0 $, and
$g_{km} = C_{kl}^p  C_{pm}^l $. Then $C_{kl}^m $ are the structure
constants of the Lie group $SL(n,{\tbf C}) $, and $g_{kl} $ its
Killing-Cartan metric tensor.
All the derivations of the algebra $M_n({\tbf C}) $ are {\it interior},
i.e. the basis of the derivations is given by the operators 
$\partial_k $ such that
\begin{equation}
\partial_k E_m = ad(i E_k) E_m = i [E_k, E_m] = C_{km}^l E_l
\end{equation}
By virtue of the Jacobi identity, we have
\begin{equation}
\partial_k \partial_m - \partial_m \partial_k = C_{km}^l \partial_l
\end{equation}
The linear space of derivations of $M_n({\tbf C}) $,  denoted by  $Der(M_n
({\tbf C}) $, {\it is not} a left module over the algebra $M_n({\tbf C}) $ 
- this is the first important difference with respect to the usual 
differential geometry, in which a vector field can be multiplied on
the left by a function, producing a new vector field.
The canonical basis of 1-forms dual to the derivations $\partial_k $
is defined formally by the relations
\begin{equation}
\theta^k (\partial_m) = \delta^k_m  {\tbf 1}
\end{equation}
These 1-forms span a left module over $M_n({\tbf C}) $, i.e. $E_l \theta^k $
is also a well-defined 1-form; indeed,
$E_l \theta^k (\partial_m) = E_l \delta_m^k {\tbf 1} = E_l \delta_m^k $ 
\newline
The exterior differential $d $ is defined as usual, first on the 0-forms
(``functions'') by the identity
\begin{equation}
df(X) = Xf ,
\end{equation}
with $f $ a function, $X $ an arbitrary vector field.
Here we have 
\begin{equation}
(d {\tbf 1}) (\partial_m) = \partial_m {\tbf 1} = ad(iE_m){\tbf 1} =
i [E_m, {\tbf 1} ] = 0
\end{equation}
so that $d {\tbf 1} = 0 $, and
\begin{equation}
d E_k (\partial_m) = \partial_m (E_k) = i[E_k,E_m] = C_{mk}^l E_l
\end{equation}
Because the Lie algebra $SL(n,{\tbf C}) $ is semi-simple, the matrices
$C_{km}^l $ are non-degenerate, and the above relation can be solved in
$\theta^k $'s giving the explicit expression
\begin{equation}
dE_k = C_{km}^l E_l \theta^m
\end{equation}
The fact that $d^2 = 0 $ follows then directly from the Jacobi identity. 
The Grassmann algebra of p-forms is defined as usual, with the wedge product 
\begin{equation}
\theta^k \wedge \theta^m = ({1\over 2}) ({\theta^k \otimes \theta^m -
\theta^m \otimes \theta^k})
\end{equation}
We have then
\begin{equation}
d{\theta^k} + ({1\over 2}) C_{ml}^k {\theta^m }{\theta^l } = 0
\end{equation}
If we define the {\it canonical 1-form} $\theta = \sum{E_k \theta^k} $,
we can easily prove that it is coordinate-independent. Moreover, it
satisfies the important relation
\begin{equation}
d{\theta} + {\theta} \wedge {\theta} = 0
\end{equation}
Let $\omega $ be a $p $ -form. The anti-derivation $i_X $  with respect to
a vector field $X $  can be defined as usual, 
\begin{equation}
(i_X \omega)(X_1,X_2, \dots ,X_{p-1}) = \omega(X,X_1,X_2, \dots ,X_{p-1})
\end{equation}
The Lie derivative of a $p $ -form $\omega $  with respect to a vector
field $X $  is defined as
\begin{equation}
{\it L}_X {\omega} = (i_X \ \ d + d\ \ i_X) {\omega} 
\end{equation}
It is easy to check now that the 2-form  $\Omega = d \theta $ is invariant
with respect to the derivations of $ A $, i.e. that
\begin{equation}
{\it L_X}{\Omega} = 0
\end{equation}
for any vector field $X $ belonging to $ Der(M_n(\tbf C)) $.
The 2-form $\Omega $ is also non-degenerate, and it is a closed 2-form
by definition, because 
\begin{equation} 
d{\Omega} = d^2 {\theta} = 0
\end{equation}
The 2-form $\Omega $ defines a Hamiltonian structure in the algebra
$M_n({\tbf C}) $ in the following sense:
\newline
\indent
Let $f \ \  \epsilon \ \  M_n({\tbf C}) $ be an element of our algebra;
then $Ham_f $  is the {\it Hamiltonian vector field of f }   defined by
the equality
\begin{equation}
\Omega (Ham_f , X) = X \ \ f 
\end{equation}
for any $X $ belonging to $Der(M_n({\tbf C})) $
The {\it Poisson bracket}  of two ``functions'' (observables) $f $ and 
$g $ is then defined as
\begin{equation}
\lbrace  f,g \rbrace = \Omega (Ham_f , Ham_g )
\end{equation}
A simple computation shows then that   
\begin{equation}
\lbrace E_k, E_m \rbrace = {\Omega} (\partial_k , \partial_m ) 
= i\  \ [E_k ,E_m]
\end{equation}
Therefore, in our simple version of non-commutative geometry, 
classical and quantum mechanics merge into one and the same structure: 
the Poisson bracket of any two matrix ``functions'' (observables) is equal, 
up to a factor that can be chosen as the Planck constant
$h $ , to their {\it commutator} .
\newline
\indent
This simple and beautiful picture is of course somewhat perturbed in 
the case of {\it infinite-dimensional algebras} for which not all the
derivations are interior and might have other representations than
the commutator with an observable.
\newline
\indent
The volume element induced by the canonical Cartan-Killing metric
and the corresponding Hodge duality $\star $ can be also introduced in a
classical manner. The volume element is given by
\begin{equation}
\eta = {1\over{(n^2 - 1)!}}\epsilon_{i_1 i_2 \dots i_{n^2 - 1}}
\theta^{i_1} \wedge \theta^{i_2} \wedge \dots  \wedge \theta^{i_{n^2-1}}
\end{equation}
Any $n^2 - 1 $ -form is proportional to the volume element $\eta $ ;
the integral of such a form will be defined as the {\it trace} of the
matrix coefficient in front of $\eta $ . Then the scalar product is
readily introduced for any couple of $p $ -forms  $\alpha $ and $\beta $
as follows:
\begin{equation}
(\alpha , \beta) = \int (\alpha \wedge \star \beta)
\end{equation}
\indent
With this formalism we can generalize the notions of gauge fields if we 
use the non-commutative matrix algebra as the analogue of the algebra of 
functions defined on a vertical space of a principal fibre bundle. Then we 
will be able to compute lagrangian densities that may be used in the 
variational principle producing dynamical field equations.
\newline
\indent
We shall see in the next section how this formulation of gauge theories
contains besides the $SU(n)$ gauge fields also scalar multiples in the
adjoint representation, which have the r\^ole of the Higgs fields in
standard electroweak theory.
\vskip 0.4cm
\indent
{\tbf 4. Non-commutative analog of Kaluza-Klein and gauge theories.}
\vskip 0.3cm
\indent
At this stage we can introduce a non-commutative analogue of Kaluza-Klein
type theory, which will lead to a generalization of gauge field theories.
In ordinary differential geometry the fact of using a Cartesian product 
of two differential manifolds, or a fibre bundle locally diffeomorphic 
with such a product, can be translated into the language of the
corresponding function algebras; as a matter of fact, in the case of
the Cartesian product of two manifolds, the algebra of functions defined
on it is the tensorial product of algebras of functions defined on each
of the manifolds separately.
\newline
\indent
Consider the space-time manifold $V_4 $ with its algebra of smooth
functions $ {\it C^{\infty}} (V_4) $. Let us define the tensor product 
\begin{equation}
{\sl A} = C^{\infty} (V_4) \otimes M_n (\tbf C)
\end{equation}
It can be shown (cf.[13]) that
\begin{equation}
{\sl Der(A) } = [{\sl Der(C^{\infty}}(V_4))\otimes {\tbf 1}] \oplus
[{\sl C^{\infty}} (V_4) \otimes {\sl Der}(M_n({\tbf C})]
\end{equation}
In other words, a general derivation in our tensor product algebra
replacing the algebra of smooth functions on a fibre bundle space,
can be written as the following vector field
\begin{equation}
X =  X^{\mu} (x) \partial_{\mu} + \xi^k (x) \partial_k
\end{equation}
with $\mu ,\nu = 0,1,2,3 $ ; $k,l = 1,2,  \dots , (n^2 - 1) $ . 
A general 1-form defined on such vector fields splits naturally into
{\it four} different components:
\begin{equation}
A = A^0_{\mu} (x) {\tbf 1} dx^{\mu} + A^k_{\mu} (x) E_k dx^{\mu} 
+ B^0_m (x) {\tbf 1} \theta^m + B^k_m (x) E_k \theta^m
\end{equation}
The exterior differential of a 1-form $A $ takes into account the two
kinds of differentiation; e.g. for a general matrix-valued 0-form 
(``function'')
$\Phi  =  \Phi^0 {\tbf 1} (x) + \Phi^m (x) E_m $ we have
\begin{equation}
d(\Phi) = (\partial_{\mu} \Phi^0 ) dx^{\mu} + (\partial_{\mu} \Phi^m) E_m
dx^{\mu} + \Phi^m C_{km}^l E_l \theta^k
\end{equation}
The notion of {\it covariant derivation} can be introduced quite
naturally by considering a free (right) hermitian module 
${\sl H} $  over the 
algebra ${\sl A } $. If we choose a {\it unitary element  e} in ${\sl H}, $ 
then any element of ${\sl H} $ can be represented as  $\Phi = e B ,$ with 
$B \in {\sl A} $. Then the {\it covariant derivative} must have 
the following basic property:
\begin{equation}
\nabla (\Phi D) = (\nabla \Phi) D + \Phi \otimes dD
\end{equation}
for arbitrary $\Phi \ \ \epsilon  \ \ {\sl H} $ ,
$D \ \ \epsilon \ \ {\sl A} $
Now, if $\Phi = eB $, we shall have
\begin{equation}
\nabla \Phi = (\nabla e) B + e \otimes dB ,
\end{equation}
and there exists a unique element $\alpha \ \ \epsilon \ \
\Lambda^1 (M_n(\tbf C)) $ such that
\begin{equation}
\nabla e = e \otimes d\alpha
\end{equation}
satisfying the hermiticity condition $\bar \alpha = - \alpha $. The elements 
$B $ and $\alpha $ are called the {\it components} of the field $\Phi $ 
and the connection $\nabla $ in the gauge $e $.
\newline
\indent
Let $U $ be a unitary matrix from the algebra $\sl A $. Under a change
of gauge
\begin{equation}
e \longrightarrow  eU
\end{equation}
the components $B $ and $\alpha $ transform as follows:
\begin{equation}
B \longrightarrow  U^{-1}B  \   \ , 
\alpha \longrightarrow U^{-1} \alpha U \ \ + \ \ U^{-1} dU
\end{equation}
This is the analogue of the gauge theory in the non-commutative case. When 
applied to the connection 1-form (denoted now $A $ instead of  $\alpha $ ),  
these principles lead to the following expression of the gauge field tensor 
$F=dA + A\wedge A :$  
$$F = (F^0_{\mu \nu} {\tbf 1} + G^k_{\mu \nu} E_k )dx^{\mu} \wedge dx^{\nu} +
[(D_{\mu} B^0_l) {\tbf 1} + (D_{\mu} B_l^m E_m)] dx^{\mu} \wedge \theta^l
+G_{kl}^m E_m \theta^k \wedge \theta^l $$
where
\begin{equation}
F^0_{\mu \nu} = \partial_{\mu} A^0_{\nu} - \partial_{\nu} A^0_{\mu}
\end{equation}
represents the abelian $U(1) $-gauge field;
\begin{equation}
G^k_{\mu \nu} = \partial_{\mu} A^k_{\nu} - \partial_{\nu} A^k_{\mu}
+ C^k_{lm} A^l_{\mu} A^m_{\nu}
\end{equation}
represents the $SU(2) $-gauge field;
\begin{equation}
D_{mu} B^0_k =({1 \over m}) ({\partial_{\mu} B^0_k})
\end{equation}
is the derivative of the scalar triplet $B^0_k $ ;
\begin{equation}
D_{\mu} B^m_k =({1\over m})({\partial_{\mu} B^m_k 
+ C^m_{sr} A^s_{\mu} B^r_k})
\end{equation}
is the covariant derivative of the scalar (Higgs type) multiplet $B^m_k $ ;
finally,
\begin{equation}
G^m_{kl} =({1\over {m^2} })(C^p_{kl} B^m_p - C^m_{sr} B^s_k B^r_l )
\end{equation}
represents the potential contribution of the Higgs multiplet.
\newline
\indent
Here $m $ is the dimensional parameter $(dim[m] = cm^{-1}) $ introduced in 
order to give the proper dimension to the 1-forms $\theta^{k}$. The parameter 
$m $ can be later related to the characteristic mass scale of the theory.
The generalized action integral is equal - in conformity with the 
definition of integration on the algebra of p-forms in the non-commutative 
case - to the {\it trace} of the integral over space-time $V_4 $
of the expression $F \wedge \star F $ :
\begin{equation}
{\tbf Tr} \int (F \wedge \star F ) d^4 x
\end{equation}
The multiplet of scalar fields $B^m_l $ plays here the r\^ole of the
symmetry-breaking Higgs-Kibble field, whose quartic potential appearing
in the last part of the action integrand possesses multiple local
minima or maxima. 
\newline
\indent
In this example, when all other fields are set equal to $0 $, there
exist several configurations of $B^m_l $ corresponding to vacuum states
representing different gauge orbits. Indeed, it is easy to see that
$G^m_{kl} = 0 $ not only when $B^m_l = 0 $, but also for 
$B^m_l = \delta^m_l $. These two vacua {\it can not } be transformed 
one into another by means of a gauge transformation, which is a novel
feature when compared with the known classical versions of gauge theory
coupled with Higgs fields.
\newline
\indent
Although this generalization of gauge theory including a non-commutative 
sector of differential geometry contains naturally the gauge group 
$SU(2) \times U(1) $, the Higgs multiplet arising here does not have 
the usually required properties, i.e. {\it it is not} a doublet of complex 
scalar fields coupled in a different way to the left- and right-handed 
fermions; we have instead a tensor multiplet $B^m_l $ that admits 16 
different vacuum configurations, most of them degenerate saddle points in 
the parameter space. Also the mass spectrum of bosons appearing in the theory 
is not satisfactory. Developing the bosonic fields of the model,
$A^0_{\mu} $,  $A^k_{\mu} $ and $B^0_k $ , and linearizing the equations
around the vacua given by $ B^m_l = 0 $ or  $B^m_l = \delta^m_l $
respectively, we obtain  on the gauge orbit $B^m_l = 0 $:
\newline
\indent
- masses of $A^0_{\mu} $ and $A^k_{\mu} $ equal zero,
\newline
\indent
- masses of $B^0_l $ and $B^m_l $ all equal to ${\sqrt n} \, m;$ 
\newline
\noindent
whereas on the gauge orbit  $B^m_l = \delta^m_l $ :
\newline
\indent
- the  $U(1) $ gauge field $A^0{\mu} $ remains massless while the $SU(2) $ -
gauge field 
\newline
\indent
acquires the mass $\sqrt{2n} \, m $  ;
\newline
\indent
- the scalar multiplet $B^0_m $ acquires the mass ${\sqrt 2} \, m $, and  
the Higgs multiplet 
\newline
\indent
itself, $B^m_l $ develops a mass spectrum with
values $0 $, ${\sqrt 2} \, m $ and $2 {\sqrt 2} \, m $ .
\newline
\noindent
which makes this version of unified $SU(2) \times U(1) $ theory unrealistic.
\newline
\indent
More realistic versions of non-commutative gauge models, reproducing quite 
well all the properties of the electroweak interactions required by the 
experiment, have been proposed by A.Connes and M.Lott [17], R. Coquereaux 
{\it et al.} [18], by M.Dubois-Violette {\it et al.}, [15], [16], and by 
J.Fr\"ohlich {\it et al.}, [19]. In all these models the non-commutative 
algebra of complex matrices is tensorized with a $Z_2 $-graded algebra, 
which in simplest realisation can be conceived as algebra of $2 \times 2 $ 
matrices that splits into two linear subspaces called {\it ``even'' } 
(corresponding to diagonal matrices) and {\it ``odd''} (corresponding to 
the off-diagonal matrices), with respective {\it grades} being $0 $ and $1 $, 
which under matrix multiplication add up {\it modulo} 2. The exterior 
derivations change the grade of an element by 1, and satisfy the 
{\it graded Leibniz rule}
\begin{equation}
d(AB) = (dA)B + (-1)^{grad(A)grad(B)} AdB
\end{equation}
This enables one to represent the connection form (interpreted as the 
gauge-field potential) in the following form:
\begin{equation}
\pmatrix{A &W^+ \cr W^- &Z}
\end{equation}
where the gauge fields $A $  and $Z $ belong to the even part of the algebra, 
while the fields $W^+ $ and $W^- $ belong to the odd part; moreover, all 
these fields are themselves $2 \times 2 $ matrix-valued 1-forms.
Developing this theory around the appropriately chosen vacuum configuration
one can quite correctly reproduce the mass spectrum, with the mass of
neutral $Z $ -boson $2\over {\sqrt 3} $ times bigger than the mass of
the charged $W $ - boson, which corresponds to the Weinberg angle of   
$30^o .$ 
More details can be found in the papers cited above.
\newline
\indent
At this point one may try to imagine what a non-commutative extension of
the General Relativity could look like ? Since a long time there exist many
approaches in which the General Relativity was considered as a gauge theory,
with gauge group being the infinite-dimensional group of diffeomorphisms
of four-dimensional Riemannian manifolds. However, with the gauge group of
this size little could be done in matter of computation and prediction,
especially on the quantum level.
\newline
\indent
A more realistic direction consists in exploring the properties of linear
approximation of a more complicated final version of the theory. Recently,
J. Madore {\it et al.} in [20] have introduced the generalization of linear
connections on matrix algebras defined above. With the usual definition of
covariant derivation acting on the moving frame:
\begin{equation}
D \, \theta^{\alpha} = - \omega^{\alpha}_{\beta} \otimes \theta^{\beta}
\end{equation}
Because the definition of covariant derivative requires that
\begin{equation}
D(f \xi) = df \otimes \xi + f \, D\xi ,
\end{equation}
the covariant derivative of an arbitrary $1$-form 
$\xi_{\alpha} \theta^{\alpha}$ is 
$$ D\, (\xi_{\alpha} \theta^{\alpha})=d \xi_{\alpha} \otimes \theta^{\alpha} 
- \xi_{\alpha} \, \omega^{\alpha}_{\, \beta} \, \theta^{\beta} $$
The covariant derivative along a vector field $X$ is defined as
\begin{equation}
D_X \, \xi = i_X \, (D \xi)
\end{equation}
and defines a mapping of $\Omega^1(V)$ on itself.
\newline
\indent
If the torsion vanishes, then one finds that
\begin{equation}
D^2 \, \theta^{\alpha} = - \, \Omega^{\alpha}_{\, \beta} \, \theta^{\beta}
\end{equation}
where $\Omega^{\alpha}_{\, \beta} = R^{\alpha}_{\, \beta \gamma \delta} \,
\theta^{\gamma} \wedge \theta^{\delta}$ is the {\it curvature 2-form}.
\newline
\indent
The generalization of these formalism for the non-commutative case is quite
obvious. We must replace the linear space of $1$-forms which span the tensor
and the exterior algebras by the corresponding right ${\cal{A}}$-module of
$1$-forms defined over our matrix algebra $\Omega^1(M_n {\bf C})).$ In the
basis introduced in the previous section, $\theta^k$, $k=1,2,...(n^2-1)$, we
had
$$d \theta^k = - \frac{1}{2} \, C^k_{\, \, lm} \theta^l \theta^m, \, \ \ \,
{\rm and} \, \ \ \, \ \ d f = [\theta, f] . $$
It is easy to define the linear connection with vanishing torsion:
\begin{equation}
D \theta^r = - \omega^r_{\, s} \otimes \theta^s \, , \, \ \ \, \ \
{\rm with} \, \ \ \, \ \ \omega^r_{\, s} = -\frac{1}{2}C^r_{\, s t} \theta^t
\end{equation}
\indent
Introducing the permutation operator $\sigma$ as
$$\sigma(\theta^{k} \otimes \theta^m) = \theta^m \otimes \theta^k ,$$
we can express the commutativity of the algebra ${\cal{C}}^{\infty}(M_4)$ 
$$D\,(\xi f) = D(f \, \xi)$$
by writing
$$D(\xi f) = \sigma(\xi \otimes df) + (D \xi) \, f .$$
\indent
The last condition can be maintained in a more general case as the requirement
imposed on the connection $1$-forms. It follows then that in the case of
matrix algebras considered here, one has
\begin{equation}
D([f, \theta^k]) = [f, D \theta^k] = 0 ,
\end{equation}
so that all the coefficients $\omega^k_{\, l m}$ must be in the center of
$M_n({\bf C})$, i.e. they are just complex numbers, and the torsionless
connection defined above becomes unique.
\newline
\indent
The metric in the space of $1$-forms over $M_n({\bf C})$ has been already
introduced as $g(\theta^k \otimes \theta^m) = g^{k m} \in {\bf C}$. The fact
that $\omega^k_{\, (l m)} = 0$ can be interpreted as the metricity of this
connection. This leads to the unique definition of the corresponding curvature
tensor:
$$\Omega^k_{\, \, l m n} = \frac{1}{8} C^k_{l r}C^r_{m n} $$
These constructions have been used already in [15] and [16], and can serve
as the non-commutative extension of connexion and curvature on the tensor
product of algebras ${\cal{C}}^{\infty}(M_4) \otimes M_n({\bf C}) .$
\newline
\indent
However, the fact that all geometrically important quantities like metric,
connection and curvature coefficients, are forced to belong to the 
{\it center} of the non-commutative sector make the above generalization
quite trivial and therefore unsatisfactory.
\vskip 0.4cm
\indent
{\tbf 5. Minkowskian space-time as a commutative limit}
\vskip 0.3cm
\indent
In this section we shall discuss an important feature of any non-commutative
geometry that contains the algebra of smooth functions on Minkowskian
space-time and is supposed to be Poincar\'e-invariant at least in the first
orders of the deformation parameter. This result has been published in 1998
(M. Dubois-Violette, J. Madore, R. Kerner, [21]). Similar ideas have been
independently developed earlier by S. Doplicher, K. Friedenhagen and 
J.E. Roberts (cf. [22]).
\newline
\indent
The main idea is as follows. Suppose that the non-commutative geometry that
is supposed to describe in an adequate way the quantum version of General
Relativity contains in its center the infinite algebra of smooth functions
on Minkowskian space-time. This infinite algebra serves as a representation
space for the infinite-dimensional representation of the Poincar\'e group,
in particular, the abelian group of translations, in the limit when the
gravitational interaction becomes negligible, which shall correspond to
the limit $\kappa \rightarrow 0$, where $\kappa$ is proportional to the
gravitational coupling constant $G$. 
\newline
\indent
It seems natural to suppose that the Poincar\'e invariance remains still
valid before the limit is attained, at least in the linear approximation 
with respect to the deformation parameter $\kappa$. Then an important question 
to be answered appears, namely, what is the dimension of the non-commutative
part of the full algebra before the limit is attained ? As it is shown in
the reference [21], it must be infinite-dimensional. In other words, it is
impossible to impose the full Poincar\'e invariance on a tensor product of
${\cal{C}}^{\infty} (M_4)$ with a {\it finite} non-commutative algebra, as 
in the example with the matrix algebras considered in previous sections. 
These examples can be considered only as approximations to the correct
theory of non-commutative space-time and gauge field theories.
\newline
\indent
Let us consider then a one-parameter family of associative algebras,
${\cal{A}}_{\kappa}$, whose limit at $\kappa = O$, denoted by ${\cal{A}}_0$,
admits a well-defined action of the Poincar\'e group on it. When 
$\kappa \rightarrow 0$, one should attain as a classical limit certain
algebra, obviously containing ${\cal{C}}^{\infty} \, (M_4)$, the algebra of 
smooth functions on the Minkowskian manifold:
\begin{equation}
{\cal{A}}_{\kappa} \rightarrow {\cal{A}}_0  \supset {\cal{C}}^{\infty} (M_4)
\end{equation}
The one-parameter family of associative algebras, ${\cal{A}}_{\kappa} ,$
can be analyzed with the help of the {\it deformation theory} developed in
the well-known article by F. Bayen, M. Flato, C. Fronsdal and A. Lichnerowicz
(cf. [23]). It is supposed that all ${\cal{A}}_{\kappa}$ coincide - as vector
spaces - with a fixed vector space $E$. The product of any two elements $f,g$
in ${\cal{A}}_{\kappa}$ can be expanded as follows:
\begin{equation}
(fg)_{\kappa} = fg + \kappa \, c(f,g) + o(\kappa^2)
\end{equation}
where $fg = (fg)_0$ is the product in ${\cal{A}}_0$. We also assume that
there is a common unit element ${\bf 1}$ for all ${\cal{A}}_{\kappa}$.
The commutators of any two elements $f, g $ in ${\cal{A}}_{\kappa}$ and in
${\cal{A}}_O$ are related via the following equation:
\begin{equation}
[f,g]_{\kappa} = [f,g]_0 - i \, \kappa \, \{ f,g \} + o(\kappa^2)
\end{equation}
where $\{ f,g \} =  i \, ( c(f,g) - c(g,f)) .$ The mapping $(f,g) \rightarrow
c(f,g)$ is called a {\it normalized Hochschild 2-cocycle} of ${\cal{A}}_0$
with values in ${\cal{A}}_0$.
\newline
\indent
The derivation property of the commutator in ${\cal{A}}_{\kappa}$ should be
maintained, which means that
\begin{equation}
[h, (fg)_{\kappa} ]_{\kappa} = ( [h,f]_{\kappa}, g)_{\kappa} + 
( f, [h,g]_{\kappa} )_{\kappa}
\end{equation}
Then, in the first order in $\kappa$, we get
\begin{equation}
i \biggl( [h, c(f,g)] - c([h,f], g) - c(f, [h,g]) \biggr) = f \{h,g \} - 
\{ h, fg \} + \{ h, f \} g
\end{equation}
\indent
This implies that if $h \in {\cal{Z}} ({\cal{A}}_0) $, the center of the
algebra ${\cal{A}}_0$, then the endomorphism $ \delta_h : \delta_h (f) =
\{ h, f \} $  is a derivation of ${\cal{A}}_0 : $
\begin{equation}
\{ h , \{ f,g \} \} = \{ \{ h,f \} , g \} + \{ f, \{ h,g\} \}
\end{equation}
The {\it center} of the algebra ${\cal{A}}_0$, denoted by 
${\cal{Z}}({\cal{A}}) ,$ is stable under these derivations, and therefore, 
it closes under the bracket $\{ \, , \, \}$. This means that the Jacobi
identity valid in all associative algebras ${\cal{A}}_{\kappa}$ remains
valid, at least up to the second order in $\kappa$, in ${\cal{A}}_0 :$
$$ {\rm from} \, \ \ \, \ \ \,
[f, [g,h]_{\kappa}]_{\kappa} + [g, [h,f]_{\kappa}]_{\kappa} +
[h, [f,g]_{\kappa}]_{\kappa} = 0   \, \ \ \, \ \ {\rm it \ \ follows \, } $$
\begin{equation}
\{ f, \{ g,h \}_{\kappa} \}_{\kappa} + \{g, \{ h,f\}_{\kappa}\}_{\kappa} +
\{ h, \{ f,g \}_{\kappa} \}_{\kappa} = 0  
\end{equation}
Summarizung up, we can make the following statement:
\newline
\indent
The center of ${\cal{A}}_0$, ${\cal{Z}}({\cal{A}}_0)$, is a {\it commutative
Poisson algebra} with the Poisson bracket given by
$$ i ( c(f,g) - c(g,f)) $$
The linear mapping $ z \rightarrow \delta_z$ maps ${\cal{Z}} ({\cal{A}}_0)$
into the Lie agebra of derivations of ${\cal{A}}_0 : \, \ \ \, \delta_z(f) =
\{ z , f \} , \, \ \ \, $ for $z, f \in {\cal{A}}_0$
\newline
\indent
We wish to represent the non-commutative analog of {\it real functions} by
Hermitian elements of the extended algebra of functions. Therefore, we should
impose the following {\it reality condition} :
\newline
\indent
\hskip 0.5cm
- all the ${\cal{A}}_{\kappa}$ are complex *-algebras, whose involutive 
vector spaces coincide with the unique space $E ;$
\newline
\indent
\hskip 0.5cm
- for any $f \in E$, also $f^* \in E$; moreover, we assume that there exists
a unique hermitian element which is the common unit for all these algebras,
${\bf 1}^* = {\bf 1}$, such that
$$(fg)^*_{\kappa} = (f^* g^*)_{\kappa} , \, \ \ \, {\rm and} \, \ \ \,
({\bf 1} f)_{\kappa} = (f {\bf 1} )_{\kappa} = f $$
\indent
It follows that the normalized co-cycle $c(f,g)$ satisfies natural condition
$$(c(f,g))^* = c(g^*, f^*)$$
\indent
Thus, the set ${\cal{Z}}_R ({\cal{A}}_0$  of all Hermitian elements of
${\cal{Z}}({\cal{A}}_0$ forms naturally a real Poisson algebra, and 
$z \rightarrow \delta_z$ maps it into the real Lie algebra $Der({\cal{A}}_0$
of all Hermitian derivations of ${\cal{A}}_0 .$
\newline
\indent
Now comes the main point: the necessary realization of the Poincar\'e 
invariance on these algebras. The family ${\cal{A}}_{\kappa}$ represents
non-commutative extensions of the algebra of smooth functions on space-time.
Even if these algebras are not Poincar\'e-invariant, we wish to recover the
Poincar\'e-invariant physics on the usual Minkowski space in the limit when
$\kappa \rightarrow 0 .$  Therefore, we must assume that the Poincar\'e
group ${\cal{P}}$ acts via *-automorphisms on the limit algebra ${\cal{A}}_0:$
\begin{equation}
(\Lambda, a) \rightarrow D_{\Lambda,a)} \in {\cal{L}} \, ({\cal{A}}_0 ,
{\cal{A}}_0 )
\end{equation}
for any element $(\Lambda, a) \in {\cal{P}}$.
\newline
\indent
By hypothesis, the algebra ${\cal{A}}_0$  contains a *-subalgebra identified
with the commutative algebra of smooth functions on Minkowski space,
${\cal{C}}^{\infty}(M_4)$. The action of ${\cal{P}}$ on ${\cal{A}}_0$ should
induce the usual action of ${\cal{P}}$ on ${\cal{C}}^{\infty} (M_4)$
associated with the corresponding linear transformations in $M_4$.
\newline
\indent
We shall now argue that ${\cal{C}}^{\infty} (M_4)$ {\it can not be the whole}
${\cal{A}}_0$.
\newline
\indent
Indeed, suppose that ${\cal{A}}_0 = {\cal{C}}^{\infty}(M_4)$. The, in view 
of the our previous satement concerning the Poisson structures, there exists
a Poisson bracket on $M_4$. This Poisson bracket must be non-trivial, since
we assumed that the ${\cal{A}}_{\kappa}$ are all non-commutative.
\newline
\indent
On the other hand, we know that there does not exist a non-trivial Poincar\'e
invariant bracket on $M_4$. Indeed, let $(f,g) \rightarrow \{f, g\}$ be
such a bracket. Then, in a given coordinate patch, it can be represented
analytically as
\begin{equation}
\{ f, g \} = \Omega^{\mu \nu} \, \partial_{\mu} f \, \partial_{\nu} g
\end{equation}
whare $\Omega^{\mu \nu} = \{ x^{\mu} \, , \, x^{\nu} \} $ must be an
anti-symmetric tensor field on $M_4$, which is constant with respect to
translations and Lorentz covariant.
\newline
\indent
However, the rotational invariance already implies that the three-vectors
$$E^i = \Omega^{0 i} \, \ \ {\rm and} \, \ \ \, B^k = \epsilon^k_{lm} \, 
\Omega^{l m}, \, \ \ \, (i, k, l = 1,2,3)$$
should vanish, which means that $\Omega^{\mu \nu} = 0$, and therefore, also
$\{ f,g \} = 0$ for all $f, g \in {\cal{C}}^{\infty} (M_4) .$
\newline
\indent
It seems unreasonable to suppose that the Poincar\'e invariance is broken at
the first order in $\kappa$, because at this order we expect to recover a
spin-2 Poincar\'e-invariant theory, coupled to other physical fields. So, if 
the Poincar\'e invariance holds at the first order in $\kappa$, it follows 
that the inclusion ${\cal{C}}^{\infty}(M_4) \subset {\cal{A}}_0$ must be a 
strict one, i.e. the limit $\kappa \rightarrow 0$ of ${\cal{A}}_{\kappa}$ 
must contain an extra factor besides ${\cal{C}}^{\infty} (M_4).$ Therefore, 
the normalized two-cocycle $c( \, , \, )$ of ${\cal{A}}_0$ defined by
\begin{equation}
(f g)_{\kappa} = fg + \kappa \, \, c(f,g) + o(\kappa^2)
\end{equation}
is supposed to be Poincar\'e-invariant, i.e. it has the property:
\begin{equation}
\alpha_{(\Lambda, a)} \, \biggl( c(f,g)) = c(\alpha_{(\Lambda,a)} (f), 
\alpha_{(\Lambda,a)} (g) \biggr)
\end{equation}
which implies the invariance of the $\kappa$-bracket:
\begin{equation}
[f,g]_{\kappa} = [f,g] - i \, \{ f,g \} + o(\kappa^2)
\end{equation}
\indent
Let us consider now the elements of ${\cal{A}}_0$ that belong to 
${\cal{C}}^{\infty} (M_4)$ and generate the commutative algebra of smooth 
functions on $M_4 : \, \ \ x^{\mu} \in {\cal{C}}^{\infty} (M_4)$. By
definition, we have then
\begin{equation}
\alpha_{(\Lambda,a)} \, x^{\mu} = {\Lambda}^{ -1 \mu}_{\, \ \ \, \nu} \, 
(x^{\nu} - {\bf 1} \, a^{\nu} )
\end{equation}
\indent
By choosing the origin, one can identify ${\cal{C}}^{\infty}(M_4)$ with the
Hopf algebra of functions on the group of translations of $M_4$. Since
${\cal{C}}^{\infty}(M_4)$ is a subalgebra of ${\cal{A}}_0$, the algebra
${\cal{A}}_O$ is a bimodule over ${\cal{C}}^{\infty}(M_4)$. As a left 
${\cal{C}}^{\infty}(M_4)$-module, ${\cal{A}}_0$ is isomorphic with the 
tensor product ${\cal{C}}^{\infty}(M_4) \otimes {\cal{A}}^I_0$ , where
${\cal{A}}^I_0$ denotes the subalgebra of transitionally invariant elements
of ${\cal{A}}_0$ :
\begin{equation}
{\cal{A}}^I_0 = \Big\{ f \in {\cal{A}}_0 \mid \alpha_{(1, a)} (f) = f \, \ \
{\rm for \ \ all} \, \ \ a \Big\}
\end{equation}
\indent
In fact, ${\cal{A}}_0$ is isomorphic with ${\cal{C}}^{\infty}(M_4) \otimes 
{\cal{A}}^I_0$ as a $({\cal{C}}^{\infty}(M_4) , {\cal{A}}^I_0)$-bimodule.
Thus in order to recover the complete algebraic structure of ${\cal{A}}_0$,
it is sufficient to describe the right multiplication by elements of
${\cal{C}}^{\infty}(M_4)$ of the elements of ${\cal{A}}^I_0$. The algebra
${\cal{A}}^I_0$ is stable under the derivations induced by the generators
of local coordinate variables $x^{\mu}$:
$$f \rightarrow ad(x^{\mu}) (f) = [ x^{\mu} , f ] $$
\indent
Therefore, for any $f \in {\cal{A}}^I_0$ one has
$$ f x^{\mu} = x^{\mu} f - ad(x^{\mu}) (f) $$
or, in the tensorial representation  ${\cal{A}}_0 = {\cal{C}}^{\infty}(M_4)
\otimes {\cal{A}}^I_0 : $
$$f x^{\mu} = x^{\mu} \otimes f - {\bf 1} \otimes ad(x^{\mu})(f) \, \ \ 
{\rm for \ \ any \, \ \ } f \in {\cal{A}}^I_0 $$
\indent
From  this we can deduce the right multiplication of ${\cal{C}}^{\infty}(M_4)
\otimes {\cal{A}}^I_0$ by the elements of ${\cal{C}}^{\infty}(M_4)$. Let us
denote by $X^{\mu}$ the four commuting derivations of ${\cal{A}}^I_0$
induced by $ad(x^{\mu})$. The algebra ${\cal{A}}^I_0$ is invariant under the 
action of the diffeomorphisms $\alpha_{(\Lambda, 0)}.$ 
\newline
\indent
Let us denote by $\alpha^I_{\Lambda}$ the homomorphism of the Lirentz group
into the group $Aut({\cal{A}}^I_0)$ of all the *-automorphisms of 
${\cal{A}}^I_0 .$
\newline
\indent
Then one can summarize the above discussion of properties of our algebra
by the following presentation of ${\cal{A}}_0 :$
\newline
\indent
We start with a unital *-algebra ${\cal{A}}^I_0$ equipped with four commuting
anti-Hermitian derivations $X^{\mu}$ and the action $\Lambda \rightarrow
\alpha^I_{\Lambda}$ of the Lorentz group through the automorphisms of
${\cal{A}}^I_0 :$
\begin{equation}
\alpha^I_{\Lambda} \circ X^{\mu} = {\Lambda}^{-1 \, \ \ \mu}_{\, \ \ \, \ \ 
\nu} \, X^{\nu} \circ \alpha^I_{\Lambda}
\end{equation}
The entire algebra ${\cal{A}}_0$ is generated as a unital *-algebra by
${\cal{A}}^I_0$ and the four Hermitian elements $x^{\mu}$ which satisfy the
relations:
$$ x^{\mu} x^{\nu} = x^{\nu} x^{\mu} $$
\begin{equation}
\, \ \ \, {\rm and} \, \ \ \, \ \ x^{\mu} f = f x^{\mu} + X^{\mu} (f) \,
\ \ \, \ \ {\rm if} \, \ \ \, \ \ f \in {\cal{A}}^I_0
\end{equation}
\indent
The Poincar\'e group acts on ${\cal{A}}_0$ as follows:
\vskip 0.2cm
\indent
\hskip 0.5cm
- for $x^{\mu} \in {\cal{C}}^{\infty}(M_4) :$
\begin{equation}
\alpha_{(\Lambda, a)} \, (x^{\mu}) = {\Lambda}^{-1 \, \mu}_{\, \ \ \, \nu}
(x^{\nu} - a^{\nu} \, {\bf 1} ) ;
\end{equation}
\indent
\hskip 0.5cm
- for $f \in {\cal{A}}^I_0 :$
\begin{equation}
\alpha_{(\Lambda,a)} (f) = \alpha^I_{\Lambda} (f) .
\end{equation}
\indent
But we have assumed before that the bracket
$$\{ f,g \} = i \, \Big( \, c(f,g) - c(g,f) \Big) $$
does not vanish identically on ${\cal{C}}^{\infty}(M_4).$ This implies that 
the functions $c^{\mu \nu}$ defined as
$$ c^{\mu \nu} = c(x^{\mu} , x^{\nu})$$
do not all vanish. On the other hand, these functions being Lorentz covariant
must belong to ${\cal{A}}^I_0$, so that we have
$$\alpha_{\Lambda,a)}(c^{\mu \nu}) = {\tilde{c}}^{\mu \nu} $$
and one has
\begin{equation}
\alpha^I_{\Lambda} \, (c^{\mu \nu}) = {\Lambda}^{-1 \, \mu}_{\, \ \ \rho}
{\Lambda}^{-1 \, \nu}_{\, \ \ \sigma} \, c^{\rho \sigma} ,
\end{equation}
so that the homomorphism of the Lorentz group into the group 
$Aut({\cal{A}}^I_0$ of the *-automorphisms of ${\cal{A}}^I_0$ is never 
trivial.
\newline
\indent
This implies in turn that ${\cal{A}}^I_0$ cannot be a finite-dimensional 
algebra (like e.g. the complex matrix algebra discussed in our previous
example), because on such an algebra all automorphisms are {\it inner}, and
on the other hand, it is known that the Lorentz group has no non-trivial,
finite dimensional unitary representations. Therefore, the extra factor
that is present in ${\cal{A}}_0$ besides the usual infinite-dimensional
algebra of functions (coordinates) on $M_4$ must be also infinite dimensional.
\newline
\indent
In view of previous analysis, the algebra ${\cal{A}}_0$ is the tensor 
product  ${\cal{C}}^{\infty}(M_4) \otimes {\cal{A}}_O^I$, with the Lorentz
group acting via automorphisms on ${\cal{A}}_0^I .$ Since the brackets
$ \{ x^{\mu}, x^{\nu} \} \in {\cal{A}}_0^I ,$ the algebra ${\cal{A}}_0^I$
must contain as a subalgebra an algebra of functions on the union of Lorentz
orbits of anti-symmetric 2-tensors. The coordinates on this algebra viewed
as a manifold are just the brackets $ \{ x^{\mu} , x^{\nu} \} .$ The orbits
may be labeled by the following two parameters:
\begin{equation}
\alpha = g_{\mu \rho} \, g_{\nu \lambda} \, \{ x^{\mu} , x^{\rho} \}
\{ x^{\nu} , x^{\lambda} \} \, \ \ \, \ \ {\rm and} \, \ \ \, \ \
\beta = \epsilon_{\mu \nu \rho \sigma} \, \{ x^{\mu} , x^{\nu} \} \, 
\{ x^{\rho} , x^{\sigma} \}.
\end{equation}
If we want to include the definitions of time reversal and parity, we should
assume that whenever a given orbit $(\alpha, \beta)$ appears in the algebra,
the orbit corresponding to $(\alpha, - \beta)$ should appear as well. When
one has also $\{ x^{\mu} , \{ x^{\nu} , x^{\lambda} \} \} = 0$ for all
values of indeces $\mu, \nu, \lambda$, then ${\cal{A}}^I_0$ is equal to
the above algebra.
\newline
\indent
The simplest situation occurs when ${\cal{C}}^{\infty}(M_4)$ belongs to the 
center of ${\cal{A}}_0$. In this case the cocycle $c$ is antisymmetric 
(up to a co-boundary) on ${\cal{C}}^{\infty}(M_4)$, and also on the center
${\cal{Z}}({\cal{A}}_0)$ itself. Then ${\cal{A}}_0$ is a commutative Poisson
algebra, and the family ${\cal{A}}_{\kappa}$ can be obtained by its
geometric quantization.
\newline
\indent
It is not difficult to give an example of such one-parameter family of
algebras, containing the usual representation of the Poincar\'e algebra
acting on smooth functions (coordinates) on $M_4$.
$$ [ x^{\mu} , x^{\nu} ] = i \kappa M^{\mu \nu}$$
$$ [x^{\lambda} , M^{\mu \nu} ] = i ( g^{\lambda \nu} L^{\mu} - 
g^{\lambda \mu} L^{\nu} ) $$
$$ [ x^{\mu} , L^{\nu} ] = i \kappa M^{\mu \nu} $$
$$ [M^{\lambda \rho} , M^{\mu \nu} ] = i (g^{\lambda \nu} M^{\mu \rho} -
g^{\rho \nu} M^{\mu \lambda} + g^{\rho \mu} M^{\nu \lambda} - g^{\lambda \mu}
M^{\nu \rho} ) $$
$$[L^{\lambda} , M^{\mu \nu}] = i ( g^{\lambda \nu} L^{\mu} - g^{\lambda \mu}
L^{\nu}) $$
\begin{equation} 
[ L^{\mu} , L^{\nu} ] = i \kappa M^{\mu \nu}
\end{equation}
where $g^{\mu \nu}$ denotes the Minkowskian metric $diag(-1, 1,1,1).$ It 
follows from the above relations that for $\kappa\neq 0$ the algebras
${\cal{A}}_{\kappa}$ are generated by the $x^{\mu}$. For any value of
$\kappa$ there exists an action of the Poincar\'e group ${\cal{P}}$ on 
${\cal{A}}_{\kappa}$ via *-automorphisms $(\Lambda, a) \rightarrow 
\alpha_{(\Lambda, a)}$ defined as:
$$ \alpha_{(\Lambda,a)} \, x^{\mu} = {\Lambda}^{-1 \, \mu}_{\, \ \ \, \nu} \, 
(x^{\nu} - a^{\nu} \, {\bf 1} ), \, \ \ \,  \alpha_{(\Lambda,a)} \, L^{\mu}
= {\Lambda}^{-1 \, \mu}_{\, \ \ \, \nu} \, L^{\nu} , \, \ \ \, 
\alpha_{(\Lambda,a)} \, I^{\mu \nu} = {\Lambda}^{-1 \, \mu}_{\, \ \ \, \rho}
\Lambda^{-1 \, \nu}_{\, \ \ \, \sigma} \, I^{\rho \sigma} .$$
\indent
The commutation relations between the $I^{\mu \nu}$ and the $L^{\lambda}$ are
the relations of the Lie algebra of $SO(4,1)$ if $\kappa$ is positive, of
$SO(3,2)$ if $\kappa$ is negative, and of the Poincar\'e algebra if $\kappa
= 0$. It follows that the $I^{\mu \nu}$ and the $L^{\lambda}$ generate
the corresponding enveloping algebras. The differences of the generators
$x^{\mu} - L^{\mu}$ are in the center $Z({\cal{A}}_{\kappa})$ of 
${\cal{A}}_{\kappa}$; therefore the algebra ${\cal{A}}_{\kappa}$ is the 
tensor product of the commutative algebra generated by the $(x^{\mu}-L^{\mu})$
and the two following Casimir operators:
$$C_2 = \kappa \, g_{\mu \nu} g_{\rho \lambda} \, I^{\mu \rho} I^{\nu \lambda}
+ 2 \, g_{\mu \nu} \, L^{\mu} L^{\nu} ,$$
\begin{equation}
C_4 = g^{\rho \rho'} \, (\epsilon_{\rho \lambda \\mu \nu} L^{\lambda}
I^{\mu \nu})(\epsilon_{\rho' \lambda' \mu' \nu'} L^{\lambda'} I^{\mu' \nu'})
\end{equation}
where $\epsilon_{\mu \nu \lambda \rho}$ is the totally anti-seymmetric tensor
with $\epsilon_{0123} = 1$. Therefore also ${\cal{A}}_0$ is the tensor product
of the commutative akgebra  generated by the $(x^{\mu} - L^{\mu})$ with the
enveloping algebra of the Poincar\'e Lie algebra generated by the $L^{\mu}$
and the $I^{\mu \nu}$.
\newline
\indent
It must be stressed here that this Poincar\'e algebra is not the same as
the Poincar\'e algebra acting on ${\cal{A}}_0$ (like on the space-time
variables) via the automorphisms $\alpha_{(\Lambda, a)}$; this can be seen
also by the fact that $L^{\mu}$ have the dimension of a length. This double
appearance of the Poincar\'e algebra may be interpreted as the necessity to
introduce matter besides the space-time itself as soon as we penetrate in the
non-commutative sector of the great algebra containing 
${\cal{C}}^{\infty}(M_4)$ as a factor.
\newline
\indent
Since our Casimirs $C_2$ and $C_4$ are contained in the center of 
${\cal{A}}_{\kappa}$, and since they are translationally invariant, we can
impose some fixed values on them, thus specifying even more precisely the
algebras ${\cal{A}}_{\kappa}$. Since the element $C_2$ has the dimension
of a length squared, and the element $C_4$ that of a length to the power
four, the most natural choices amount to attribute the value $\kappa^2$ 
to the element $C_4$, while the element $C_2$ can be given the following 
three particular values:
$$i) \, C_2 = \kappa ; \, \ \ \, \ \  ii) \,  C_2 = - \kappa , \, \ \ \, \ \
iii) \, C_2 = 0 . $$
\indent
All these choices lead to $g_{\mu \nu} L^{\mu} L^{\nu} =0$ in ${\cal{A}}^I_0.$
Remembering the fact that ${\cal{A}}^I_0$ has the structure of the enveloping
algebra of the Poincar\'e Lie algebra, the last condition is an analogue of 
the {\it zero mass} condition in the ususal case.
\newline
\indent
With the value of $C_4$ fixed in such a way that the representations found
here are all of ``zero mass'' and ``strictly positive spin'' type, which
gives the algebra ${\cal{A}}^I_0$ a characteristic two-sheet structure,
corresponding to the two possible helicities, which in turn results from
the fact that the Lorentz group is not simply connected.
\newline
\indent
As a concluding remark, we would like to stress the fact that in general the
Poincar\'e covariance of ${\cal{A}}_{\kappa}$ is not necessary; all we need
here is to ensure the Poincar\'e covariance of ${\cal{A}}_0$ only. Another
deformation of the Poincar\'e algebra, called ``the $\kappa$-Poincar\'e''
has been studied in a series of papers published recently by J. Lukierski and 
co-authors ([24]).
\newline
\indent
Their approach is in some sense complementary to the scheme presented above:
instead of considering the action of the {\it exact} Poincar\'e group on 
the space-time containing a non-trivial deformation because of the supposed
non-commutative character of the coordinates, one chooses to consider the
action of a {\it deformed} Poincar\'e group, called the $\kappa$-Poincar\'e,
on the ordinary space-time. It seems plausible that in the linear limit
both these approaches nearly coincide.
\newpage
\indent
{\tbf 6. Quantum Spaces and Quantum Groups}
\vskip 0.3cm
A more radical deformation of usual behaviour of functions describing
the coordinates and their differentials consists in modifying the
commutation relations not only between the coordinates and their
differentials, but also between the coordinates themselves, and between
the differentials as well, which would represent a very profound modification 
of the space-time structure. Moreover, if we look for the transformations
that would keep these new relations invariant, we discover that such
transformations can not be described by means of ordinary groups, which
therefore need to be generalized. Such new generalizations have been
introduced by V.Drinfeld, L.Faddeev and S.L.Woronowicz, ([25], [26], [27]).
and they are known under the name of "Quantum Groups". 
\newline
\indent
The litterature on this subject is very abundant; we shall cite the papers by
S.L.Woronowicz [27], as well as the papers of L.C.Biedenharn [28], J.Wess 
and B.Zumino [29], L.A.Takhtajan [30], V.G.Kac [31]; the list is far from 
being exhaustive, so that we shall limit ourselves to an outline of the 
main idea illustrated by a simple example.
\newline
\indent
Conformally with the spirit of quantum field theories, the most important
mathematical object to be studied is the algebra of observables, which are
usually functions of few fundamental ones. This approach can be extended
to the mathematical study of Lie groups: indeed, we can learn almost
everything concerning group's structure from the algebraic structure of 
functions (real or complex) defined on it.
\newline
\indent
Consider a compact manifold $G $ which is also a Lie group; let $e $  denote 
its unit element. The algebra $\sl{A} $ of functions defined on $G $ has a 
very particular structure, which is implemented by the following three 
mappings:
\medskip
$i) $ for each $f\epsilon {\sl A} $, there is an element of ${\sl A}\otimes 
{\sl A} $, denoted by $\Delta f $, such that $\Delta f (x,y) = f(xy) $;
The mapping  $\Delta $ : 
\begin{equation}
{\sl A} \longrightarrow {\sl A \otimes \sl A } 
\end{equation}
is called the {\it{coproduct}}.
\medskip
$ii) $ There exists a natural mapping from $\sl A $ into {\tbf C} 
(or {\tbf R}) defined by
\begin{equation}
\epsilon : f \longrightarrow  f(e) \, \in \, {\tbf C}
\end{equation}
which is called the {\it{co-unit}}
\medskip
$iii) $ There exists a natural mapping of ${\sl A} $ into itself:
\begin{equation}
(Sf)(x) = f(x^{-1} ) 
\end{equation}
which is called the {\it antipode }
\newline
\indent
It is easy to see that in the case of the algebra of functions defined
on a Lie group the co-product is non-commutative if the Lie group is
non-commutative; however, the multiplication law in the algebra $\sl A $
itself remains commutative as long as we consider the functions taking their
values in {\tbf C} or {\tbf R}.
This particular structure of an associative commutative algebra $\sl A $
with the three operations defined above, the {\it co-product}, the
{\it co-unit} and the {\it antipode} is called the {\it Hopf algebra}.
Now, the natural extension that comes to mind is to abandon the postulate
of the commutativity of the product in $\sl A $; in this case, the 
structure is named the {\it Quantum Group}. It should be stressed that 
a quantum group is not a group, but a general algebra which only in the
commutative case behaves as the algebra of functions defined on a Lie
group.
\newline
\indent
One of the most interesting aspects of this theory is the fact that the 
quantum goups arise quite naturally as the transformations of non-commutative 
geometries known under the name of {\it quantum spaces} introduced by 
Yu.Manin, J.Wess and B.Zumino, and others. We shall illustrate how a quantum 
group can be constructed on a simple example in two dimensions called 
the {\it Manin plane} ([32]).
\newline
\indent
Consider two ``coordinates'' $x $ and $y $ spanning a linear space
and satisfying
\begin{equation}
xy \ \ = q\ \ yx 
\end{equation}
with a complex parameter  q different from 1.
Consider a transformation
\begin{equation}
x' = a\ \ x + b\ \ y ,   \ \   y' = c\ \ x + d\ \ y .
\end{equation}
which preserves the relation $xy $ = q $yx $, i.e. such that
\begin{equation}
x'y' = q\ \ y'x'
\end{equation}
We shall suppose that the quantities $a,b,c,d $ commute with the 
``coordinates'' $x,y $; the simplest realization of this requirement
is achieved by assuming (disregarding the nature of the entries of the
matrix) that the multiplication of $x $ by $a,b, $ etc. is tensorial,
i.e. when we set by definition 
\begin{equation}
x' = a\otimes x  + b\otimes y .
\end{equation}
Then the conservation of the q-commutation relations between $x $ and $y $
leads to the following rules for $a,b,c $ and $d $:
\begin{equation}
ac = q\ \ ca, \  \ bd = q\ \ db, \  \ ad = da + q\ \ cb - ({1\over q}) bc
\end{equation}
In order to fix all possible binary relations between the coefiicients
$a,b,c $ and $d $ we need three extra relations, which would define
$bc $, $ab $ and $cd $. Such relations can be obtained if we define the
``differentials''
\begin{equation}
\xi  = dx , \  \  \eta  = dy , \  \  \xi^2 = 0, \  \  \eta^2 = 0.
\end{equation}
satisfying twisted $p $ -commutation relations
\begin{equation}
\xi \eta + ({1\over p}) \eta \xi = 0
\end{equation}
with a new complex parameter $p $.
Assuming that the exterior differentiation commutes with the transformation
matrix and requiring the same relations for $\xi ' $ and $\eta ' $, we get
\begin{equation}
bc = ({q\over p})cb,\ \ ab = p\ \ ba,\ \ cd = p\ \ dc .
\end{equation}
With these relations the matrix algebra defined above becomes associative
and can be given the structure of a Hopf algebra as follows:
$$\Delta \pmatrix{a &b\cr c &d } = \pmatrix{a\otimes a + b\otimes c
&a\otimes b + b\otimes d\cr c\otimes a + d\otimes c &c\otimes b +
d\otimes d}$$
and the co-unit as
$$\epsilon \pmatrix{a &b\cr c &d} = \pmatrix{1 &0\cr 0 &1} $$
The antipode $S $ of a quantum matrix should be defined as its inverse.
In order to make such a definition operational, we need a non-commutative
generalization of the determinant of a matrix. Such a ``($q,p $ )-determinant''
should be defined as the combination of parameters appearing in the
transformation law for the ``elementary area element'', i.e. the exterior
product of the differentials $\xi $ and $\eta $:
\begin{equation} 
\xi' \eta' = {D_q} \xi \eta
\end{equation} 
which yields immediately
\begin{equation}
D_q = ad - p bc \ \ = da - ({1\over q})bc
\end{equation}
The determinant $D_q $ commutes with $a $ and $d $, but has non-trivial
commutation relations with the off-diagonal elements $a $ and $b $ 
(in what follows, we shall omit the subscript $q $ for the sake of 
simplicity) :
\begin{equation}
Db = ({p\over q})bD, \ \ Dc = ({q\over p})cD .
\end{equation}
It should also possess an inverse $D^{-1} $, which in fact is a new
element extending the algebra, and satisfying
\begin{equation}
D^{-1}D = \tbf{1} = DD^{-1}
\end{equation}
Applying these identities to the commutation relations verified by $D $,
one finds easily that $D^{-1} $ commutes with $a $ and $b $, and 
satisfies 
\begin{equation}
bD^{-1} = ({p\over q})D^{-1} b,\ \ cD^{-1} = ({q\over p})D^{-1} c
\end{equation}
It is easy to see that 
\begin{equation}
\Delta (D) = D\otimes D,\ \ \Delta (D) \Delta (D^{-1}) = \Delta (\tbf{1})=
\tbf{1} \otimes \tbf{1}
\end{equation}
and
\begin{equation}
\Delta (D^{-1}) = D^{-1} \otimes D^{-1}
\end{equation}
The antipode of any matrix can be determined now as follows:
\begin{equation}
S\pmatrix{\pmatrix{a &b \cr c &d}}= D^{-1} \pmatrix{d &({-1\over q})b\cr
-qc &a} = \pmatrix{d &({1\over p})b \cr -pc &a}D^{-1}
\end{equation}
Also
\begin{equation}
S(D)=D^{-1}, \ \ S(D^{-1}) = S(D)
\end{equation}
but $S^2  \not = \tbf{1} $. The inverse of the antipode mapping can be
also defined as
\begin{equation}
S\pmatrix{\pmatrix{a &b \cr c &d}} = D\pmatrix{a &pqb \cr ({1\over {pq}})
c &d} 
\end{equation}
\indent
The algebra generated by the matrices defined above, whose entries $a,b,c $
and $d $ satisfy the $(q,p) $-commutation relations is a Hopf algebra; it                   
is denoted by $ GL_{p,q} (2, {\bf C}).$ 
\newline
A differential calculus on such algebras has been developed by 
S.L.Woronowicz; the notion of covariant differentiation, if it can be
introduced properly, may lead to new and rich extensions of the ideas of 
connections, curvatures and gauge fields. Here we shall give an example of 
the realization of covariant derivation and the curvature $2$-form on the 
quantum plane introduced above. These results belong to M. Dubois-Violette 
{\it et al.}, published in ([33]). 
\newline
\indent
The algebra of forms on the quantum plane is generated by four elements,
$x, y, \xi = dx $ and $\eta = dy$, with the following commutation relations:
$$xy = q \, yx,$$
$$x \xi = q^2 \xi x , \, \ \ \, \ \ x \eta = q \, \eta x + (q^2-1) \, \xi y ,
\, \ \ \, \ \ y \xi = q \, \xi y, \, \ \ \, \ \ y \eta = q^2 \, \eta y ,$$
$$\xi^2 = 0, \, \ \ \,  \eta^2 = 0, \, \ \ \,  \eta \xi + q \xi \eta = 0$$
where $q$ is supposed not to be a root of unity. In (still hypothetical !) 
future physcical applications the value of the parameter $q$ is supposed 
to be very close to $1$, and in the linear approximation can be written as
$1 + \kappa$.
The above conditions are of course compatible with the definitions 
$\xi = d x, \, \eta = d y$ and with the Leibniz rule, i.e. if we apply the 
operation $d$ to the first constitutive identity $x y = q \, yx$, we obtain 
a relation which is a direct consequence of the four constitutive relations 
between $x, \, y$ and their differentials $\xi, \, \eta$ , and so forth.
\newline
\indent
All the relations between the variables $x, \, y$ and their differentials
$\xi, \, \eta$ can be written in a more uniform way using a matrix notation
which introduces the tensorial product of linear spaces spanned by both $x, y$
and $\xi, \eta$ variables. Denoting $x$ and $y$ by $x^i$ and $\xi$ and $\eta$
by $\xi^k$, with $i, k = 1,2$, we can write
$$ x^i x^j - q^{-1} {\hat{R}}^{ij}_{\, \, kl} \, x^k x^l ,$$
$$ x^i \xi^j - q \, {\hat{R}}^{ij}_{\, \, kl} \, \xi^k x^l , $$
\begin{equation}
\xi^i \xi^j + q \, {\hat{R}}^{ij}_{\, \, kl} \, \xi^k \xi^l .
\end{equation}
The tensor product of two $2$-dimensional spaces is $4$-dimensional, but the
indices that are grouped two by two can be re-labeled with their values 
ranging from $1$ to $4$, and the $R$-matrix can be written as an ordinary
$4 \times 4$ matrix:
\begin{equation}
\hat{R} = \pmatrix{q & 0&0&0 \cr 0 &(q-q^{-1}) & 1&0 \cr 0&1&0&0 \cr
0&0&0&q}
\end{equation}
If the $SL_q(2, {\bf C})$ matrix (corresponding to the case $p=q^{-1}$ in 
the more general notation  $GL_q(p,q) (2, {\bf C})$ introduced above) is
written, with the same indices $k,l = 1,2$ as
$$ a^i_{\, k} = \pmatrix{a&b \cr c&d} $$
then the invariance of the $q$-commutation relations with respect to the
simultaneous transformation of the linear spaces $x, y$ and $\xi, \eta$ by
a matrix belonging to the quantum group $SL_q(2, {\bf C})$ amounts to the
following relation:
\begin{equation}
{\hat{R}}^{ij}_{\, \, kl} \, a^k_{\, m} \, a^l_{\, n} = a^i_{\, k} a^j_{\, l}
\, {\hat{R}}^{kl}_{\, \, mn}
\end{equation}
If we extend trivially the action of the differential $d$ onto the quantum
group $SL_q(2, {\bf C})$ itself by requiring all the coefficients $a^i_{\, k}$
to be constant,
$$ d \, a^i_{\, k} = 0 ,$$
\indent
The coaction of $SL_q(2,{\bf C})$ on the $x^i$ and the $\xi^k$ can be
defined then as follows:
\begin{equation}
{\tilde{x}}^i = a^i_{\, k} \otimes x^k , \, \ \ \, \ \ {\tilde{\xi}}^j =
a^j_{\, m} \otimes \xi^m .
\end{equation}
It can be found without much pain that the new variables ${\tilde{x}}^i$ and
${\tilde{\xi}}^k$ satisfy the same twisted commutation relations as formerly
$x^i$ and $\xi^k .$
\newline
\indent
As in the case of the matrix model of non-commutative geometry, one can
introduce a {\it canonical} $1$-form by defining
$$ \theta = x \, \eta - q \, y \, \xi ,  \, \ \ \, {\rm satisfying} \,  \ \ 
\, \ \  \theta^2 = 0 .$$
and is invariant under the coaction of $SL_q(2,{\bf C})$ with ${\tilde{\theta}}
= {\bf 1} \otimes \theta$ and has the following commutation relations with
the variables $x^k, \xi^m :$
\begin{equation}
x^k \, \theta = q \, \theta x^k ; \, \ \ \, \ \ \xi^m \, \theta = - q^3  \,
\theta \, \xi^m
\end{equation}
\indent
Up to a complex multiplicative constant this is the unique element of 
$\Omega^1$ (the space of $q$-one forms) verifying the above properties.
\newline
\indent
To define covariant derivation, we must introduce first the permutation 
operator $\sigma$ mapping the tensor product $\Omega^1 {\otimes}_{\cal{A}} 
\Omega^1$ into itself. As a matter of fact, the operator $\sigma$ turns out
to be just the inverse of the matrix $q \, {\tilde{R}}^{ij}_{\, \, kl} .$
We can write it down using the explicit indices $i,j,..$ as follows:
$$ \sigma \, ( \xi \otimes \xi) = q^{-2} \, \xi \otimes \xi, \, \ \ \, \ \ 
\sigma\, (\xi \otimes \eta) = q^{-1} \, \eta \otimes \xi, $$
\begin{equation}
\sigma \, (\eta \otimes \xi) = q^{-1} \, \xi \otimes \eta - (1-q^{-2}) \, 
\eta \otimes \xi , \, \ \ \, \ \ \, \sigma \, (\eta \otimes \eta) = q^{-2} \,
\eta \otimes \eta
\end{equation}
as well as 
$$\sigma \, (\xi  \otimes \theta) = q^{-3} \, \theta \otimes \xi , \, \ \  
\, \ \ \sigma \, (\theta \otimes \xi) = q \, \xi \otimes \theta - (1-q^{-1})   
\, \theta \otimes \xi ,$$
\begin{equation}
\sigma \, (\eta \otimes \theta) = q^{-3} \, \theta \eta , \, \ \ \, \ \ 
\sigma \, (\theta \otimes \eta ) = q \, \eta \otimes \theta - (1-q^{-2}) \,
\theta \otimes \eta .
\end{equation}
and also 
$$ \sigma \, (\theta \otimes \theta ) = q^{-2} \theta \otimes \theta $$
\indent
If we suppose that $q^2 \neq -1$, then the exterior algebra is obtained by
dividing the tensor algebra over $\Omega^1$ by the ideal generated by the
three eigenvectors :
$$ \xi \otimes \xi, \, \ \ \, \eta \otimes \eta \, \ \, \ \, {\rm and} \, \ \ 
\, \ \ \eta \otimes \xi + q \, \xi \otimes \eta ,$$
corresponding to the eigenvalue $q^{-2}.$ 
\newline
\indent
The {\it symmetric} algebra of forms is obtained by dividing the tensor
algebra over $\Omega^1$ by the ideal generated by the eigenvector $\xi \otimes
\eta - q \, \eta \otimes \xi$ corresponding to the eigenvalue $-1 .$
\newline
\indent
There is a unique one-parameter family of covariant derivatives compatible
with the algebraic structure of the algebra of forms defined above. It is
given by
\begin{equation}
D \, \xi^k = l^{-4} \, x^k \, \theta \otimes \theta
\end{equation}
where the parameter $l$ must have the dimension of a length. From the 
invariance of $\theta$ it follows that $D$ is invariant under the coaction
of $SL_q(2, {\bf C})$. The analog of torsion vanishes identically.
\newline
\indent
Finally, the analog of the curvature tensor can be defined here as
\begin{equation}
D^2 \, \xi^k = \Omega^k \otimes \theta = - \Omega^k_{\, j} \otimes \xi^l
\end{equation}
with the curvature $2$-forms given by the following matrix:
\begin{equation}
\Omega^i_{\, j} = l^{-4} \, (1+q^{-2})(1+q^{-4}) \, \pmatrix{ q^2 \, x y &
-q \, x^2 \cr q^2 \, y^2 & - x y } \, \xi \, \eta
\end{equation}
\indent
It vanishes for the particular values of $q$, namely, when $q = \pm i$
or $q^2 = \pm i$, but is different from zero when $q = 1 .$ The Bianchi
identity is trivially satisfied. 
\newline
\indent
No metric structure compatible with this structure can be introduced except
for the trivial case when $q = 1 .$
\vskip 0.4cm
\indent
{\tbf 7. Conclusion}
\vskip 0.3cm
\indent
We tried to present here a few versions of non-commutative generalizations
of differential geometry which are believed to serve - hopefully in some
foreseable future - as new mathematical tools that will help us to describe 
the effects of quantum gravity. Frankly speaking, in spite of beauty and
sophistication of certain models, it is hard to share this belief.
\newline
\indent
It does not mean that our efforts should be reduced or stopped at once.
``{\it Ars longa, vita brevis}'' , and there is still a lot of time ahead,
especially as compared to the cosmological scale. The overall impression
might be pessimistic, but there is always plenty of things to do.
\newline
\indent
For example, if we look at the diagram of Sect.1, we can note that besides
the {\it ``Relativistic Quantum Field Theory''} there is another unexplored
corner, the {\it ``Non-Relativistic Quantum Gravity''}. Maybe we should pay
some more attention to this direction, too ? Or at least, if such a theory
can not be formulated, try to give valuable reasons why this is the unique
combination of limits of fundamental constants that can not be realized as
a coherent theory ?

\newpage

\vskip 1cm
\noindent
{\tbf References}

\bigskip

\indent
1. J.E. Moyal, {\it Quantum mechanics as a statistical theory}, Proc.
\newline    
\indent
\hskip 0.5cm
Camb.Phil.Soc., {\tbf 45}, 99 (1949)
\vskip 0.25cm
\indent
2. P.A.M. Dirac, {\it The Fundamental Equations of Quantum Mechanics},
\newline
\indent   
\hskip 0.5cm
Proc.Roy.Soc., {\tbf A 109}, p.642 (1926); also {\it On Quantum Algebras}, 
\newline
\indent   
\hskip 0.5cm
Proc. Cambridge Phil.Soc., {\tbf 23} p.412 (1926)
\vskip 0.25cm
\indent
3. P.A.M. Dirac, as cited in {\it The Mathematical Intelligencer} {\tbf 11}
\newline   
\indent
\hskip 0.5cm
p.58 (1989)
\vskip 0.25cm
\indent
4. R. Arnowitt, S. Deser, C. Misner, in {\it Recent Developments in
\newline    
\indent
\hskip 0.5cm
General Relativity}, ed. L. Witten, Wiley, London- New York, (1962)
\vskip 0.25cm
\indent
5. B. DeWitt, Physical Review, {\bf 160}, p.1113 (1963)
\vskip 0.25cm
\indent
6. J.M. Souriau, {\it Structure des Syst\`emes Dynamiques}, Dunod, Paris, (1969)
\vskip 0.25cm
\indent
7. D. Simms, in {\it Differential Geometrical Methors in Math. Physics},
\newline
\indent
\hskip 0.5cm
Springer Lecture Notes in Mathematics, {\bf 170}, ed. K. Bleuler and A.Reetz,
\newline
\indent
\hskip 0.5cm
Springer-Verlag, (1970).
\vskip 0.25cm
\indent
8. B. Kostant, {\it Quantization and Unitary Representations}, Lecture Notes
\newline
\indent
\hskip 0.5cm
in Mathematics, {\bf 170}, Springer Verlag p.237 (1970).
\vskip 0.25cm
\indent
9. J. von Neumann, {\it Mathematical Foundations of Quantum Mechanics},
\newline
\indent
\hskip 0.5cm
Princeton University Press, (1955)
\vskip 0.25cm
\indent
10. A. Ashtekar, in {\it Mathematical Physics towards the 21-st century},
\newline
\indent
\hskip 0.7cm
R. Sen and A. Gersten eds., Ben Gurion University of the Neguev Press,
\newline
\indent
\hskip 0.7cm
p 230 (1994).
\vskip 0.25cm
\indent
11. C. Rovelli, Nucl. Physics B {\bf 405}, p.797 (1993), also the lectures
\newline
\indent
\hskip 0.7cm
in this volume.
\vskip 0.25cm
\indent
12. D. Quillen, {\it Topology} {\bf 24}, p. 89 (1985)
\vskip 0.25cm
\indent
13. A. Connes, {\it Non-Commutative Geometry}, Acad. Press., New York (1994)
\vskip 0.25cm
\indent
14. M. Dubois-Violette, C.R.Acad.Sci. Paris, {\bf 307}, S\'er.I, p.403 (1989)
\vskip 0.25cm
\indent
15. M. Dubois-Violette, R. Kerner and J. Madore, Journ. of Math. Physics,
\newline
\indent
\hskip 0.7cm
{\bf 31} (2), p. 316 (1990) 
\vskip 0.25cm
\indent
16. M. Dubois-Violette, R. Kerner and J. Madore, {\it ibid}, p. 323 (1990)
\vskip 0.25cm
\indent
17. A. Connes and J. Lott, Nucl. Phys. Proc. Suppl. B {\bf 18}, p.29 (1990)
\vskip 0.25cm
\indent
18. R. Coquereaux, G. Esposito-Far\`ese and J. Vaillant, Nucl. Phys B,
\newline
\indent
\hskip 0.7cm
{\bf 353}, p.689 (1991)
\vskip 0.25cm
\indent
19. A. Chamseddine, G. Felder and J. Fr\"olich, Commun. Math. Physics,
\newline
\indent
\hskip 0.7cm
{\bf 155}, p. 205 (1993)
\vskip 0.25cm
\indent
20. J. Madore, T. Masson and J. Mourad, Class. and Quantum Gravity, {\bf 12},
\newline
\indent
\hskip 0.7cm
p. 1429 (1995)
\vskip 0.25cm
\indent
21. M. Dubois-Violette, J. Madore and R. Kerner, Journal of Math. Physics,
\newline
\indent
\hskip 0.7cm
{\bf 39}, No 2, p. 730 (1998)
\vskip 0.25cm
\indent
22. S. Doplicher, K. Friedenhagen and J.E. Roberts, Comm. Math. Phys., 
\newline
\indent
\hskip 0.7cm
{\bf 172}, p. 187 (1995).
\vskip 0.25cm
\indent
23. F. Bayen, M. Flato, C. Fronsdal, A. Lichnerowicz and D. Sternheimer,
\newline
\indent
\hskip 0.7cm
Ann. Phys. (N.Y.) {\bf 111} p. 61, p. 111 (1978)
\vskip 0.25cm
\indent
24. J. Lukierski, A. Nowicki and H. Ruegg, Phys. Lett. B {\bf 293}, p.419 
(1992)
\vskip 0.25cm
\indent
25. V.G. Drinfeld, {\it Quantum Groups} , in Proceedings of Intern. Congress 
\newline    
\indent
\hskip 0.7cm
of Mathematics, Berkeley, p.798
\vskip 0.25cm
\indent
26. L.D. Faddeev, Proc. of Schladming Winter School, p.89 (1989)
\vskip 0.25cm
\indent
27. S.L. Woronowicz, Communications in Math. Phys. {\bf 111} p. 613 (1987)
\newline
\indent
\hskip 0.7cm
also: P. Podle\'s and S.L. Woronowicz, {\it ibid}, {\bf 178}, p.61 (1996)
\vskip 0.25cm
\indent
28. L. Biedenharn, {\it Quantum Groups}, Lecture Notes in Physics, {\bf 370},
\newline
\indent
\hskip 0.7cm
H.D. Doebner and J.D. Hennig eds., Springer-Verlag, p.67 (1990)
\vskip 0.25cm
\indent
29. J. Wess and B. Zumino, Nucl. Phys. B (Proc. Suppl.) {\bf 18}, p.32 (1990)
\vskip 0.25cm
\indent
30. L.A. Takhtajan, Adv. Studies in Pure and Applied Mathematics, 
\newline
\indent
\hskip 0.7cm
{\bf 19}, p.435 (1989)
\vskip 0.25cm
\indent
31. V.G. Kac, {\it Colloque Dixmier}, p. 471, Progrss in Mathematics, Vol. 92
\newline
\indent
\hskip 0.7cm
Birkh\"auser, Boston (1990); see also : C. de Concini and V.G. Kac, 
\newline
\indent
\hskip 0.7cm
J. Am. Math. Soc. {\bf 5} (1), p. 151 (1992).
\vskip 0.25cm
\indent
32. Yu. Manin, Commun. Math. Physics, {\bf 123}, p.163 (1989)
\vskip 0.25cm
\indent
33. M. Dubois-Violette, J. Madore, T. Masson and J. Mourad, Lett. 
\newline
\indent
\hskip 0.7cm
in Math.Physics, {\bf 35}, p. 351 (1995).

\end{document}